# Nb$_3$Sn wire Shape and Cross Sectional Area Inhomogeneity in Rutherford Cables

U. Kelly, S. Richter, C. Redenbach, K. Schladitz, C. Scheuerlein, F. Wolf, P. Ebermann, F. Lackner, D. Schoerling, D. Meinel

*Abstract*— During Rutherford cable production the wires are plastically deformed and their initially round shape is distorted. Using X-ray absorption tomography we have determined the 3D shape of an unreacted Nb$_3$Sn 11 T dipole Rutherford cable, and of a reacted and impregnated Nb$_3$Sn cable double stack. State-of-the-art image processing was applied to correct for tomographic artefacts caused by the large cable aspect ratio, for the segmentation of the individual wires and subelement bundles inside the wires, and for the calculation of the wire cross sectional area and shape variations. The 11 T dipole cable cross section oscillates by 2% with a frequency of 1.24 mm (1/80 of the transposition pitch length of the 40 wire cable). A comparatively stronger cross sectional area variation is observed in the individual wires at the thin edge of the keystoned cable where the wire aspect ratio is largest.

*Index Terms*—Rutherford cable, Nb$_3$Sn, quantitative, tomography, image processing

## I. Introduction

R UTHERFORD CABLES made of initially round wires are commonly used as conductors for accelerator magnet coils. The use of such high current cables is required to keep the coil inductance in acceptable limits. Due to their flat shape they are convenient to wind. Moreover they can be produced with the tight dimensional tolerances required for high magnet field quality. Rutherford cables provide a high packing factor needed for high engineering critical current densities, and the wires are transposed in order to reduce losses. Keystoned Rutherford cables make it easier to give the coil the desired quasi-circular geometry around the aperture.

As an example, more than 7000 km of keystoned Rutherford cable made of Nb-Ti wires have been used for the coils of the LHC magnets [1,2]. The magnets for the LHC High Luminosity upgrade (HL-LHC) [3] and the Future Circular Collider (FCC) study are produced using Rutherford type cables made of Nb$_3$Sn superconducting wires. Bi 2212 high temperature superconducting round wires can also be used for the production of Rutherford cables [4].

Rutherford cables are produced by applying compressive stress on the cable faces using four cylindrical rollers, which give the cable the required dimensions [5,6]. The strong plastic wire deformation encountered during the cabling process compacts the cable sufficiently such that it is mechanically stable and can be used subsequently for coil winding [7]. The plastic deformation of the wires is strongest at the cable edges, and in keystoned cables the compaction is particularly severe at the thin cable edge.

The cabling process can influence the Cu stabilizer RRR, as well as its electromechanical properties. A critical current reduction is observed in the cabled wires with respect to the virgin round wires, even for Nb-Ti wires [1], whose superconducting properties are not strongly strain sensitive.

Rutherford cables can be characterized using metallographic cross sections. In such cross sections the deformation of the wires and their subelements can be visualized in great detail. However, a 3D method such as X-ray micro tomography (µ-CT) [8] or neutron tomography [9] is needed to provide a full quantitative description of the cable and wire geometry. The spatial resolution that can be achieved with these techniques is typically 1/1000 of the required field of view, i.e. in the order of 1 µm for a superconducting wire, and in the order of 10 µm for a Rutherford cable.

In this study we have used X-ray tomography combined with sophisticated image processing in order to obtain a complete quantitative description of the 3D geometry of an unreacted Nb$_3$Sn keystoned Rutherford cable, a double cable stack after reaction and impregnation, and the same cable in a 11 T dipole magnet coil. The effect of a transverse pressure of 230 MPa on the outer cable geometry has been verified with the impregnated

U. Kelly, S. Richter, C. Redenbach, and K. Schladitz are with Fraunhofer Institute for Industrial Mathematics (ITWM), Kaiserslautern, Germany.

C. Scheuerlein, F. Wolf, P. Ebermann, F. Lackner and D. Schoerling are with European Organization for Nuclear Research (CERN), CH-1211 Geneva 23, Switzerland, (corresponding author phone: ++41 (0)22 767 8022, e-mail Christian.Scheuerlein@cern.ch).

D. Meinel is with Federal Institute for Materials Research and Testing (BAM), Berlin, Germany.

41LP1-0121LP1-01                                                                                               2
double cable stack. For the first time we provide a quantitative description of the cabling effect on the cable cross sectional area variation.

## II. Experimental

### A. The cable samples

11 T dipole [10] $Nb_3Sn$ Rutherford cables made of RRP type wires [11] have been characterized. Out of the 40 wires with a nominal diameter of 0.7 mm in the cable, 12 are RRP 132/169 wires that contain 132 superconducting subelements, and 28 are RRP 108/127 wires (108 subelements). The 11 T dipole cable [12] with a nominal width of 14.7 mm, a mid-thickness of 1.25 mm, a keystone angle of 0.79° and a transposition pitch of 100 mm has a 25 µm-thick stainless steel core in order to increase the contact resistance between opposing wires. The cable packing factor, defined as the ratio of the cross sectional areas of all wires in the cable and the cable width × mid-thickness is 87%.

Tomograms of an unreacted cable without insulation and impregnation and a reacted and impregnated double cable stack have been characterized. The double stack is made of two reacted cables that are stacked onto each other, with the thin edge of one cable on the thick edge of the other cable in order to obtain two parallel cable surfaces. The cables are insulated with S2-glass fiber, and after the reaction heat treatment impregnated with CTD-101K from Composite Technology Development [13].

The double cable stack has been used for mechanical transverse compression tests at ambient temperature [14,15]. The tomogram of the double cable stack has been acquired in a cable region showing half the indent with a transverse pressure of 230 MPa-tomogram and half of the tomogram is outside the indent (Fig. 1).

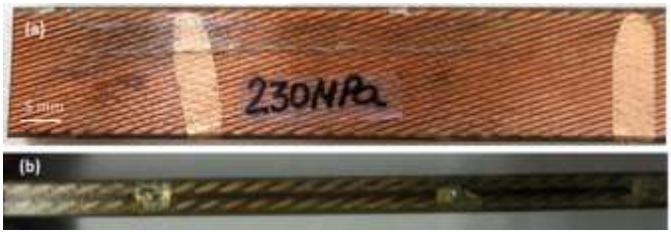

Fig. 1. Photograph of reacted and impregnated double cable stack view on (a) large surface and (b) on small surface. Note the crack in the epoxy resin after application of a transverse compressive stress of 230 MPa.

### B. X-ray tomography

The X-ray µ-CT examinations were done by the Federal Institute for Materials Research and Testing (BAM), Berlin. In order to achieve a high-resolution image, a µ-CT scanner with a 225 kV microfocus X-ray tube with 6 µm focal spot size and a flat panel detector (2048 × 2048 pixel) with a detector pixel pitch of 0.2 mm were used. The beam energy was 215 kV and 0.5 mm Cu + 0.5 mm Ag pre-filters were used. The beam current was 75 µA. The 3D volume reconstruction was based on a cone-beam algorithm of Feldkamp [16]. The voxel edge sizes in the tomograms of the reacted double cable stack and the unreacted cable are 10.5 µm and 8.8 µm, respectively. The X-ray µ-CT of a 2 cm-thick 11 T dipole segment has been performed at BAM with a beam energy of 290 kV and a voxel edge size of 25.4 µm.

### C. Image processing

Image processing at the Fraunhofer Institute for Industrial Mathematics (ITWM) was performed with the software MAVI (Modular Algorithms for Volume Images) and the modular software ToolIP (Tool for Image Processing). MAVI [17] was used to create volume renderings of 3D images to visualize the cables. The ToolIP [18] algorithms were used to segment the images for quantitative analysis of wire and cable deformation and cross sections.

The large cable aspect ratio causes artefacts in the tomographic reconstructions were corrected using a mask image. The most important steps in constructing this mask image were to first split the image into two parts - one containing the first cable and the other containing the second cable. Then both images were padded by adding 100 columns to the left and right side of the image, and 100 rows to the top and bottom of the image. Subsequently a morphological closure with a cuboidal structuring element of size $61 \times 61 \times 9$ pixels was applied. Finally, applying a global threshold returns a binary image that is a rough outline of the cable. The mask is constructed by removing the padding from both images and appending the halves.

By blurring the inverted image of this mask (using a Gauss filter) and subtracting it from the original image the brightness can be corrected. To complete the segmentation of the image, the fact that the wire Cu centers are easily segmented is useful. By labeling each wire center (i.e. one label for each wire), the image can be partitioned into segments, where each segment contains exactly one wire. By doing so, each wire can be distinguished from any neighboring wires. In Fig. 2(a) the result of the wire segmentation from the background is overlayed on the original reconstructed µ-CT slice, showing that the outer wire contours are reasonably well reproduced in the binary images (Fig. 2(b)) used for quantitative analysis of wire shape and dimensions.

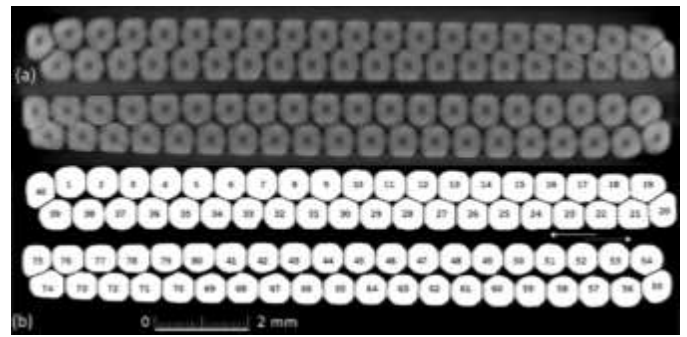

Fig. 2. (a) Visualization of the segmentation result with respect to reconstructed µ-CT image of reacted 11 T $Nb_3Sn$ double cable stack and (b) binary image of the same slice.

### III. RESULTS

*A. Wire deformation and cross sectional area variations in the unreacted 11 T dipole Nb₃Sn cable*

A reconstructed µ-CT slice of the unreacted 11 T Nb₃Sn cable and the same µ-CT slice after segmentation of the individual wires from the background are shown in Fig. 3.

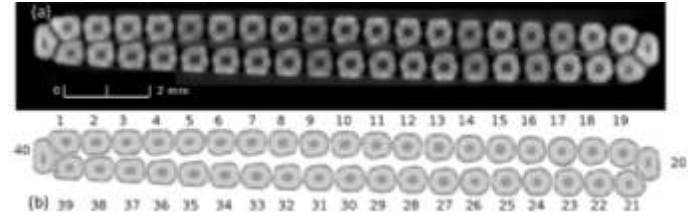

Fig. 3. (a) Reconstructed µ-CT slice of unreacted 11 T Nb₃Sn cable made out of Ø=0.7 mm RRP 108/127 and RRP 132/169 wires. (b) The same µ-CT slice after wire segmentation from background and of the subelement bundles from the Cu stabilizer. A movie showing the sequence of all µ-CT slices is available ( http://cds.cern.ch/record/2275236?ln=en ).

The 3D rendering is shown in Fig. 4. The outer Cu stabilizer has been transparently depicted on the wires 18, 19 and 20 on the thick cable edge, exposing the subelement bundles.

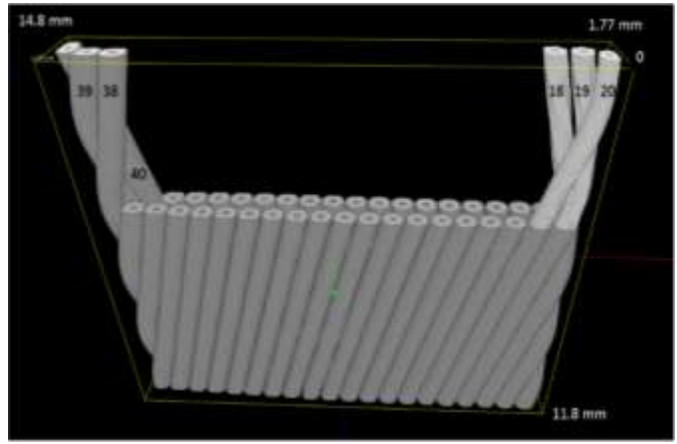

Fig. 4. Volume rendering of the unreacted 11 T dipole cable. Only wires 38, 39 and 40 are shown in their entirety. Wires 18, 19 and 20 are shown without their copper stabilizer, exposing the subelement bundles.

The distortion of the initially round wires during cabling has been quantified by measuring the aspect ratio of each wire in each of the 1421 µ-CT slices. The aspect ratio variation of the RRP 132/169 wires No. 17 and No. 39 is presented in Fig. 5. The wire and subelement bundle deformation is most severe at the thin cable edge with a maximum aspect ratio of about 1.6 and 2.1, respectively.

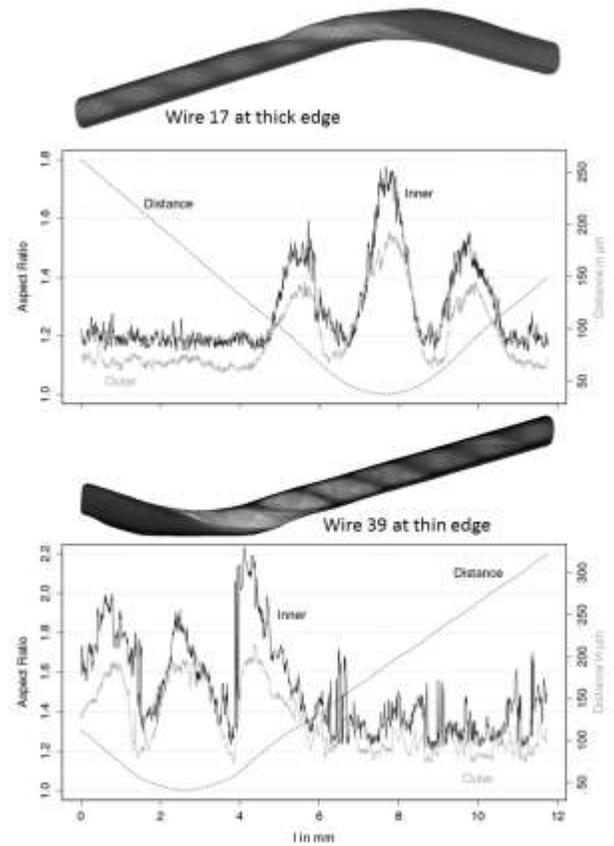

Fig. 5. Aspect ratio variation of wires 17 and 39 along the cable length $l$. The aspect ratio of the entire wire (outer) is compared with that of the subelement bundle (inner). The dotted lines indicate the wire center distance to the cable edge.

Since the cable is produced from two different wires, the µ-CT image analysis is an opportunity to compare the cabling effect on shape and cross sectional area variations of two wires with different layout. In Fig. 6 the average wire cross sectional area variation of both wires across the cable width is compared.

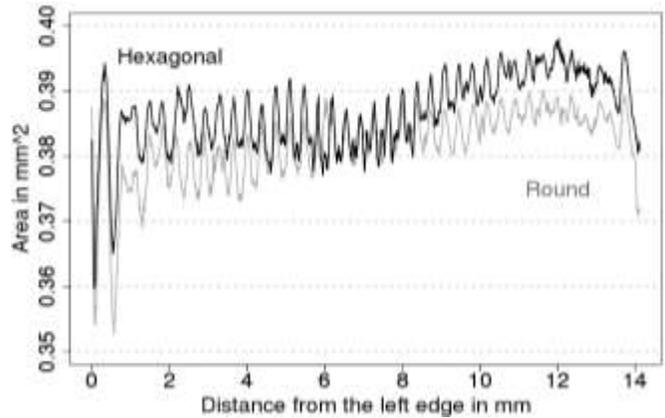

Fig. 6: Average cross sectional area of all RRP 108/127 (hexagonal filament array) and all RRP RRP 132/169 ("round" filament array) wires as a function of the wire center distance from the thin cable edge.

The RRP 132/169 wire has a slightly smaller cross section than wire RRP 108/127. The respective average values determined by digital image analysis are 0.3818 mm² and 0.3868 mm², respectively.

The cross sectional area variations of the wires across the cable width are caused by wire length changes during the cabling process. They are strongest at the thin cable edge where the wire deformation is most severe, as revealed by the aspect ratio measurements shown in Fig. 5. The effect of wire compaction on cross sectional area appears to be similar for both wires. No significant difference between the aspect ratio of both wires was observed.

*B. Cable cross sectional area variations*

The sum of the cross sectional areas of all wires in the unreacted 11 T dipole $Nb_3Sn$ cable at a given longitudinal cable position is shown in Fig. 7 over the cable length $l$. The average cable cross section determined by image analysis of the 1421 µ-CT slices is 15.39±0.085 mm$^2$, which is 3.7% smaller than the cross section of 15.99 mm$^2$ calculated for the non-reacted 11 T cable with the nominal cable dimensions. This difference between the nominal cable dimensions and the tomographic analysis is probably attributed to uncertainties in the voxel edge size and the segmentation process. However, the relative variations shown here are not strongly influenced by such uncertainties.

The cable cross sectional area oscillates by about 2% with a frequency of $l$=1.24 mm, which is one quarter of the cable length after which each wire is transposed to the adjacent position (the transposition pitch of the 40 wire cable determined from Fig. 7 is 98.9 mm).

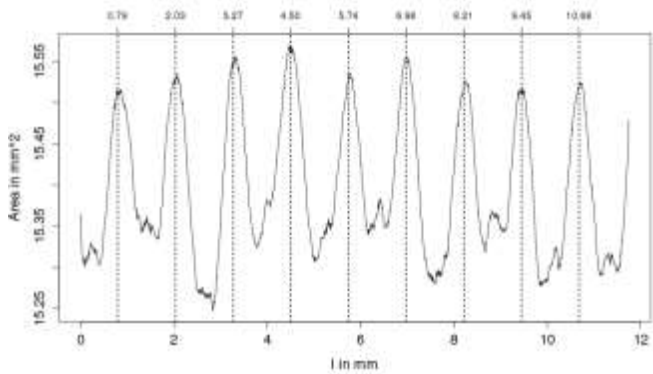

Fig. 7. Cable cross-sectional area variations along the cable length.

The corresponding µ-CT slices at which the cross sectional minima and maxima are obtained are presented in Fig. 8.

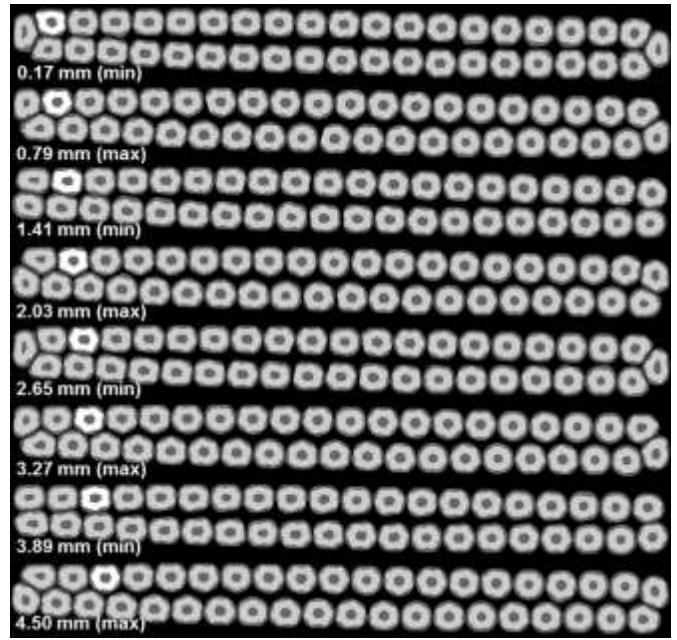

Fig. 8: Tomographic slices where the minimum and maximum cable cross sectional areas are measured. One wire is highlighted to illustrate the wire tilt in the cable.

*C. Wire deformation and cross sectional area variations in the reacted $Nb_3Sn$ 11 T dipole cable double stack*

A reconstructed µ-CT slice of the reacted $Nb_3Sn$ cable double stack, and the binary image of the same slice after segmentation of the individual wires from the background are shown in Fig. 2 (a) and (b). The spatial resolution and signal to noise ratio of the double stack cable tomogram do not allow to characterize the subelement arrays, but a segmentation of the outer wire contours is possible.

A 3D view of the two cables is presented in Fig. 9. An imprint from the transverse compressive load applied with a steel die that has visibly degraded the epoxy impregnation of the double stack cable (Fig. 1) is not seen in the cable surface rendering, indicating that plastic cable deformation under the transverse compressive load of 230 MPa was not sufficiently strong that it could be measured by the µ-CT experiment.

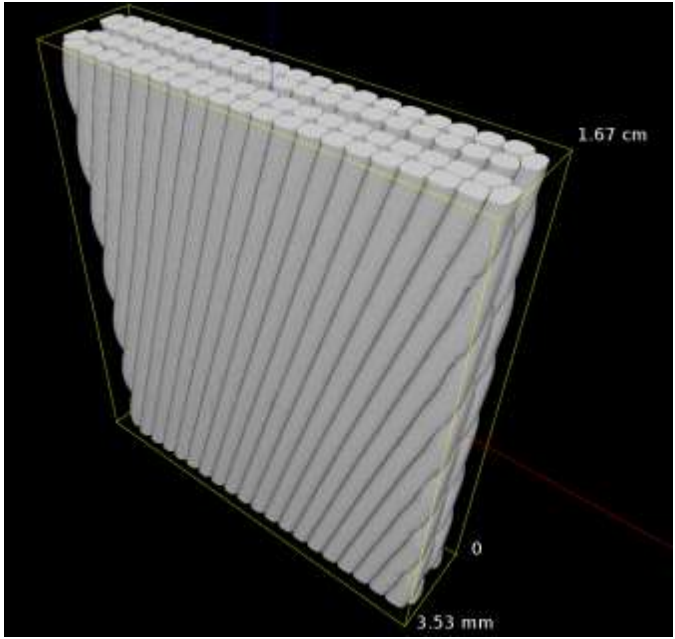

Fig. 9. 3D rendering of reacted double cable stack (wires 1 to 80).

In Fig. 10 the average thickness and wire aspect ratio of cable 1 (wires 1 to 40) is presented as a function of the distance from the thick cable edge. The average thickness of the wires in the cable (without the inter-wire void space that is filled with epoxy) is 1.054±0.208 mm.

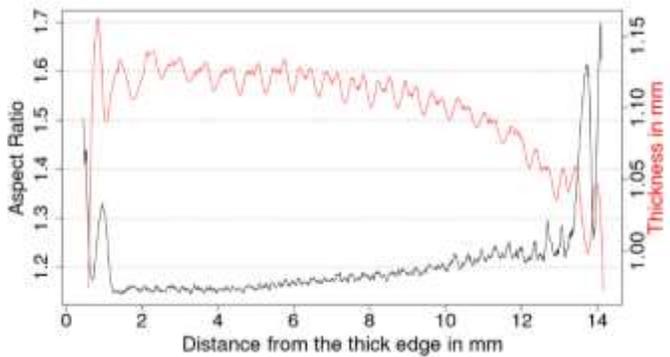

Fig 10. Average cable thickness without inter wire void space and aspect ratio of cable 1 (wires 1-40) plotted against the distance from the thick edge.

### D. Cable thickness and width in a 11 T dipole coil

The thickness of the thick and thin cable edge in a reacted and impregnated 11 T dipole coil segment has been measured by digital image analysis of µ-CT cross sections (Fig. 11). In the reacted coil the measured cable width is 14.76 mm and the mid-thickness is 1.296 mm, which is 0.4% and 3.6% larger than the respective nominal dimensions of the unreacted cable. The thin edge and thick edge thickness values measured in the µ-CT slice in the centre of a 2 cm-thick segment are 1.192±0.027 mm and 1.399±0.037 mm, respectively.

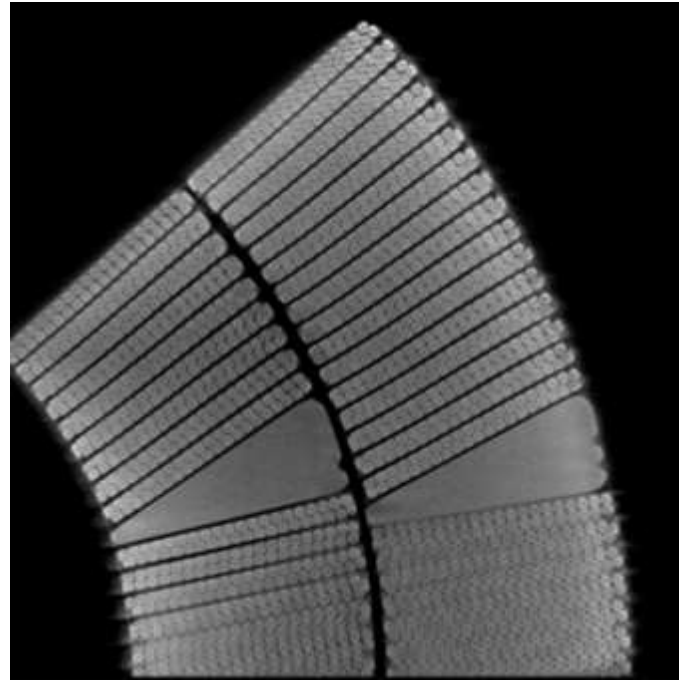

Fig 11. µ-CT slice of a 11 T dipole coil segment.

The main results describing the cable geometry are presented in Table I.

TABLE I
COMPARISON OF MAIN CABLE GEOMETRY RESULTS IN THE IMPREGNATED DOUBLE STACK AND IN A 11 T DIPOLE COIL. *WITHOUT INTERWIRE VOID SPACE. **INCLUDING VOID SPACE

|  | Mean | Thick edge | Thin edge |
|---|---|---|---|
| Wire aspect ratio | 1.206±0.082 | ~1.5 | ~1.7 |
| Wire cross section (mm2) | 0.382±0.005 | 0.385 | 0.368 |
| Cable thickness (mm)* | 1.054±0.208 | ~1.16 | ~1.00 |
| Cable thickness in coil (mm)** | n.m. | 1.399±0.037 | 1.192±0.027 |

## IV. DISCUSSION

Beyond the visualization of the cable cross sections, the combination of state-of-the-art µ-CT with sophisticated image processing treatments has enabled a quantitative analysis of the Rutherford cable geometry. With the three-dimensional tomographic reconstructions at hand the distortion of the individual wires and cross sectional area variations could be quantified.

To our knowledge we have reported here for the first time the amplitude of the wire and cable cross sectional area oscillations. The 11 T dipole cable cross sectional area oscillates by about 2% with a frequency of 1/80 of the 40 wire Rutherford cable transposition pitch length. This cross sectional area variations explain at least partly the cabling induced critical current and n-value degradation that is observed even in Nb-Ti cables that are not strain sensitive. The relatively small wire cross section at the thin cable edge may also contribute to the reduced thermal stability at the thin edge of Rutherford cables [9].

We did not observe a strong influence of the wire architecture (hexagonal subelement array in RRP 108/127 vs. "round" subelement array in RRP 132/169) on the wire and subelement array deformation during cabling. The spatial resolution of the tomography experiment is insufficient to study the deformation of individual subelements.

An imprint from the die that applied a transverse compressive stress of 230 MPa at ambient temperature to half of the double cable stack that was examined could not be revealed by the μ-CT experiment.

The cable geometry influences the cable stiffness under transverse compressive stress, and the 3 D wire and cable shape can be used as input for FE simulations [14] in order to predict electromechanical behavior of Rutherford cables under application of transverse stress in magnet coils.

ACKNOWLEDGMENT

This work was supported by the European Commission under the FP7 project HiLumi LHC under Grant GA 284404, co-funded by the DoE, USA and KEK, Japan.